# Decentralised Governance for Autonomous Cyber-Physical Systems

Kelsie Nabben[1*[0000-0003-4684-4113]] and Hongyang Wang[2*[0000-0003-4544-0715]], Michael Zargham[3[0000-0001-5279-690X]]

[1] European University Institute, 50014 Fiesole FI, IT
[2] ETH Zurich, 8092 Zurich, CH
[2] Block Science, Oakland, CA, US
```
*kelsie.nabben@eui.eu and *wang@ibi.baug.ethz.ch, both authors
                    contributed equally to this paper
```

**Abstract.** This paper examines the potential for Cyber-Physical Systems (CPS) to be governed in a decentralised manner, whereby blockchain-based infrastructure facilitates the communication between digital and physical domains through self-governing and self-organising principles. Decentralised governance paradigms that integrate computation in physical domains (such as 'Decentralised Autonomous Organisations' (DAOs)) represent a novel approach to autonomous governance and operations. These have been described as akin to cybernetic systems. Through the lens of a case study of an autonomous cabin called "no1s1" which demonstrates self-ownership via blockchain-based control and feedback loops, this research explores the potential for blockchain infrastructure to be utilised in the management of physical systems. By highlighting the considerations and challenges of decentralised governance in managing autonomous physical spaces, the study reveals that autonomy in the governance of autonomous CPS is not merely a technological feat but also involves a complex mesh of functional and social dynamics. These findings underscore the importance of developing continuous feedback loops and adaptive governance frameworks within decentralised CPS to address both expected and emergent challenges. This investigation contributes to the fields of infrastructure studies and Cyber-Physical Systems engineering. It also contributes to the discourse on decentralised governance and autonomous management of physical spaces by offering both practical insights and providing a framework for future research.

## 1    Introduction

Public blockchain infrastructure is dependent on the governance of physical components of the technology stack. For example, Bitcoin validators, as a core component of the function and governance of public blockchain networks, have always been concerned with mining hardware and related energy requirements that are required to participate in the system [1]. Blockchain projects are increasingly cognisant of the decentralised, physical infrastructure networks that are essential to the operation of decentralised technology ecosystems (known as "DePIN") [2]. Yet, conceptual frameworks for the decentralised governance of physical infrastructure networks that effectively bridge the digital-physical divide in the context of blockchain technology and autonomous systems are lacking. It is important to consider the design principles of Cyber-Physical Systems (CPS), before highly scaled, decentralised physical infrastructure network systems can be effectively deployed and governed. As such, this paper responds to the research question: 'Can physical spaces be managed through decentralised governance to achieve autonomous self-agency?.' Drawing on the field of cybernetics and an empirical case study of an autonomous house called 'no1s1', we present foundational principles and considerations for the decentralised governance of physical infrastructure networks.

"Cyber-Physical Systems", as an application of cybernetics, refers to integrations of computation, networking, and physical processes creating a symbiotic relationship between the cyber (computational) and physical (real-world) contexts [3]. These systems typically involve embedded, computational networks, consisting of interconnected networks of sensors, actuators, and computational devices, all working in tandem to monitor, control, and optimise physical processes. Circular information flows and feedback loops establish simple feedback and control loops, consisting of "observability" (the ability to infer knowledge from outputs) and "controllability" (the ability to steer the system, including physical processes, using a control input) to monitor and control the system towards self regulation [4]. These computational principles can also be applied to physical systems. The fundamental concepts of CPS are critical when it comes to thinking about technology, automation, and society at various scales. CPSs are pervasive in modern society, found in various domains such as manufacturing, transportation, healthcare, and smart infrastructure. They enable automation, real-time monitoring, and adaptive responses to changing environmental conditions, leading to improved efficiency, safety, and reliability. Some researchers refer to the integration of cyberspace, physical space, and social space as "Cyber-Physical Social Systems" (CPSSs), highlighting the interplay between physical, virtual, and social worlds, as well as the need to ensure that such complex systems are governed or "controlled" [5–7].

Conceptually, there are numerous dimensions to navigate in terms of what constitutes decentralised and autonomous governance in relation to physical infrastructure networks, and in relation to blockchain-based systems. Rather than solely focusing on DAO governance in the narrow sense of a blockchain-based system of rules for self-governance [8], blockchain-based DAOs can been observed as a form of Cyber-Physical System [19]. Indeed, the phrase "Decentralized Autonomous Organization" was coined by computer scientist and cybernetician, Werner Dilger, in 1997 to describe an intelligent home system [9]. Motivated by Dilger's conception of Decentralised Autonomous Organisation for smart homes, researchers have explored blockchain-based collective governance of their shared physical spaces via experimentation with controlling the lightness and hue of a lamp (called "LampDAO") [10].

In cybernetic terms, decentralisation refers to distribution of political power and control of an object, rather than physical distribution of the object itself. As pioneering cybernetician Stafford Beer states, "No viable organism is either centralized or decentralized. It is both things at once, in different dimensions" [11]. Relatedly, the concept of autonomy can be extended beyond basic interpretation as freedom from external political influence. In the book "Anarchist Cybernetics", Swann defines autonomy as functional, relating to the internal operation of a system, drawn from engineering "autonomous" systems, as well as political [12]. This framing can be further broken down into strategic autonomy (how decisions are made), and tactical autonomy (how organizational functions are executed) [12, 13].

## 2 Methodology

This paper considers the necessary cybernetic concepts for the decentralised governance of Cyber-Physical Systems. To do so, we adopt a qualitative case study approach, to apply the theoretical principles of CPS to the decentralised governance of autonomous physical infrastructure. Case studies allow for the study of complex phenomena in context, and are often used for the inductive exploration of yet unknown phenomena, such as theory generation [14], [15]. The case study we introduce is that of an autonomous house, called no1s1 [16], to explore the various governance considerations and limitations of facilitating decentralisation and autonomy in physical infrastructure network governance. In this instance, the insights from the study of no1s1 allow us to generate and outline a framework for decentralised governance of autonomous physical infrastructure.

This study advances understandings of decentralised governance in CPS by analysing no1s1 within the context of cybernetics. It elucidates the practical considerations and challenges of decentralised management in autonomous physical spaces, offering a foundation for future research and development in this rapidly evolving field. The remainder of this paper is structured as follows: we first explore the foundational principles of CPSs. We then examine a practical implementation of a CPS in the no1s1 case study experiment, analysing it within the context of CPS. Subsequent sections delve into considerations, challenges, and limitations in the decentralised governance of CPS. Finally, we synthesise the findings and implications of the study, including highlighting further research directions.

## 3 The Principles of CPS

Cybernetics refers to "the science of effective organization", in order to establish viable organisational structures that are capable of producing themselves [11]. Viable and sustainable self-governance can be understood as "the ability of an organizational system to sustainably implement its purpose, while maintaining homeostasis in its interaction with its niche" [17]. The practice of cybernetics lays out a broad approach that is grounded in engineering methods to governance as "steering" systems of information [18]. In this context, systems are viewed as not static and linear, but as dynamic and complex. In systems theory, open systems are viewed as interrelated components that are kept in a state of dynamic equilibrium by feedback loops of information and control [19]. Information flows are based on actuators (levers of control), sensors (detectors or signal providers), and feedback loops, that assess the system output, to enable the system to consider its performance and make adjustments to reach the desired output response [20].

CPS' are viewed as an application of cybernetics. The key architectural components of CPS' include sensors, actuators, control systems, and communication infrastructure. What these components establish are the set of parameters through which an organisation's rules may be modified. Together the mutable parameters of the organisation and of the tunable components of the technical infrastructure establish the boundaries for control of the system, (known as the "governance surface") [18].

Feedback loops are essential to CPSs as information processing systems. The purpose of feedback loops is circularity of information, for continual self-learning as humans and machines self-organise and self-govern via digital and physical infrastructures [21]. This maintenance quality is a key component of resilient infrastructure, as a system must leverage feedback loops in order to "learn to learn", and continually re-evaluate its policy and strategy against its purpose and culture (known as "recursive governance") [17]. Feedback channels in traditional engineering often consist of audits and performance assessments, reporting systems, and incident analysis. According to systems dynamics theory, the complex, behavioural dynamics of a system arise from two types of feedback loops: positive (reinforcing) and negative (balancing) [19, 22]. Degradation of the reinforcing loops over time, which are representative of the safety control structure, would inevitably lead to an accident. However, negative balancing loops, including regulation and oversight, allow people to monitor, react, and control (or 'steer') changes.

In contrast, feedback channels in blockchain networks differ from traditional systems, due the decentralised nature of the system. In addition to oversight and feedback loops that are provided via centralised regulatory bodies, rules in blockchain networks are prescribed in the software code of the protocol andincentives and/or penalties that reinforce or punish certain behaviours. Yet, there is a lack of legibility and monitoring in blockchain protocols due to the lack of central authority and the distributed organisational nature of decentralised technology projects. This can result in ineffective feedback loops in terms of the efficacy and resilience of certain behaviours, and difficulty steering the system. Ineffective feedback loops lead to vulnerabilities which can emerge across multiple dimensions of a decentralised system (including social, technical, economic, and legal [39]). Given the fairly nascent nature of cybernetic governance logic in blockchain-based contexts, combined with the prevalence of decentralised physical infrastructure networks, clear frameworks are needed to help guide effective implementation of cybernetic governance principles to decentralised CPSs. The section that follows analyses how the principles of cybernetics and Cyber-Physical Systems (CPS) apply in decentralised governance by mapping the governance surface of a case study of an autonomous cabin, called no1s.

## 4 Case Study: No1s1

No1s1 is an example of a CPS. As shown in Appendix 1, no1s1 (no-one's-one) is designed to be a smart meditation cabin that is capable of holding its own funds and eventually participating in the P2P crypto-economy as an CPS agent that utilises blockchain technology. The no1s1 prototype was designed by a researcher at ETH Zurich, and built with the support of colleagues. It was tested under various real-life scenarios across multiple live experiments and shows including locations such as World Economy Forum ETH pavilion, ETH Student Project house and University of Zurich. The

no1s1 prototype aims to prove the technical feasibility of a self-owning house, as well as providing a testing ground for further research on its transformative potential for decentralised CPS governance [16].

The statement, "A house is a machine for living in", by modernist architect Le Corbusier underscores the idea that a house can be viewed as a machine that is functional, efficient, and responsive to the needs of its occupants [23–25]. This idea hints to the integrated computation networks and automation may enhance the functionality of a house in relation to its purpose. With the advancements of smart and intelligent building technology, spaces are becoming more equipped with integrated sensors, actuators, and interconnected systems, enabling adaptation and optimisation of operations in real-time [26–28]. Intelligent spaces can monitor energy usage, adjust climate control, and manage security systems autonomously and in response to user preferences, leading to improved functionality, convenience, and sustainability. In the common applications of CPS in intelligent buildings, the controllers of the financial system are only humans, and very often centralised to landlords and/or their real-estate agencies. In blockchain-based systems, governance and operations are decentralised.

Parallel to the development of intelligent physical spaces, blockchain technology is undergoing greater integration with hardware. One notable trend is DEPIN, where physical infrastructures are decentralised through cryptocurrency-based incentive mechanisms (known as "cryptoeconomics") [2]. Another is Nature 2.0, that combines ideas from blockchain technology, Internet of Things, Artificial Intelligence, and decentralised governance to create self-sustaining, autonomous systems that mimic natural ecosystems [29, 30]. The core idea is to leverage advanced technologies to create systems that can operate independently, self-regulate, and potentially self-repair without human intervention, much like natural ecosystems do. Another experimental example of this is "Plantoid", a metal flower with blockchain enabled self-agency [31]. By leveraging blockchain-based smart contracts, it is capable of interacting and partnering with humans to collect funds and self-replicate.

Combining the above theses, the no1s1 experiment extends the logic of smart homes from machine automation to machine self-agency and human-machine collaborative governance, leveraging the ethos of engineered ownership [32] and DAO [33]. Engineered ownership highlights the possibility that machines can self-own cryptocurrency wallet addresses and store funds, interacting with human agents on blockchain network as an autonomous agent, extending the concept of cyber-focused programmable ownership in crypto-economy to CPS-focused engineered ownership. Similarly, the interaction between self-owned machine agents and human actors outside cyber spaces can be contemplated with the theories of DAO. A decentralised governance approach can be explored to assist the interaction between human and machine agents on-chain.

This makes no1s1, the intelligent cabin, a case whereby the CPS has a level of self-ownership of its own treasury and can direct interaction with human actors.

## 5   No1s1 Feedback Loops

The no1s1 prototype aims to demonstrate the technical feasibility of a smart building functioning as an autonomous Cyber-Physical System (CPS) agent with self-controlled finance, thereby enabling future decentralised governance experiments on top of it. The fundamental concept behind designing no1s1 as a CPS agent revolves around enabling financial self-agency, allowing it to self-determine rental prices for its space, as well as operational self-agency, enabling access control to users. In terms of infrastructure, no1s1 consists of a front-end user interface, a physical wooden structure, a back-end Raspberry Pi controller, an electronic system, and blockchain-based smart contracts responsible for managing payments, storing states, and enforcing access rules. The control of a cryptocurrency wallet address is delegated to the back-end controller of the cabin, eliminating the need for human intervention in accessibility decisions. The self-sustaining solar-battery energy system then determines the "life" and "death" ("on" and "off") of the system, governing the system's operational status, to essentially dictate its availability for rental and influencing its pricing to users. The data and logic in blockchain-based smart contracts serve as the authoritative source of truth for all crucial information of no1s1, such as energy levels and occupancy states of the house.

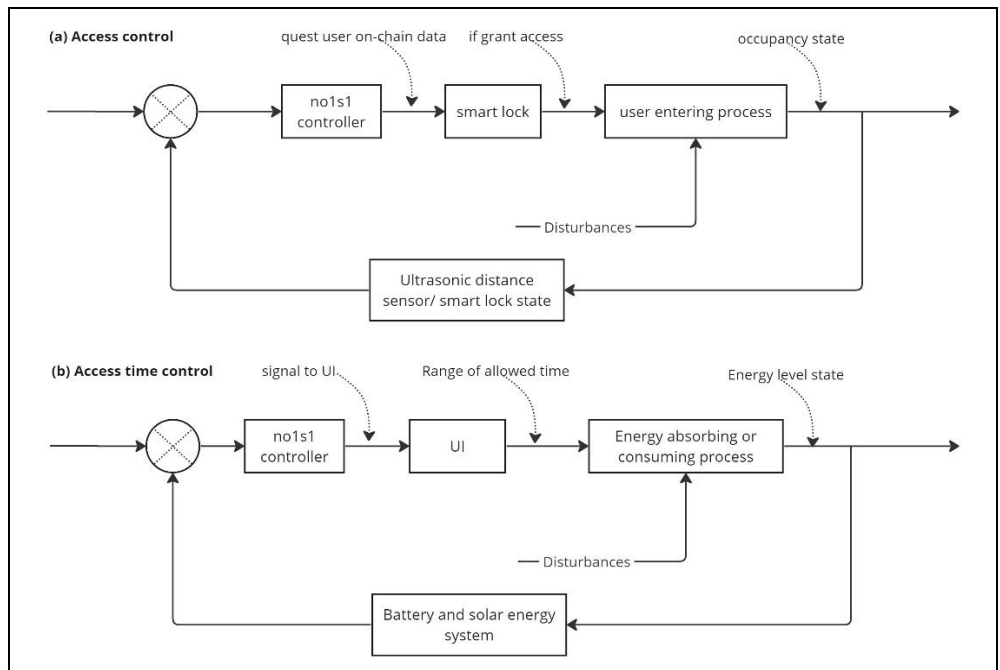

**Fig. 1.** no1s1 feedback loops

There are two primary feedback loop sets that control the access experience, as illustrated in Fig. 1. The first logic loop set regulates the rentable timeframe for no1s1, as depicted in Figure 1(b). The maximum and minimum rentable durations for renters are determined based on the energy level signal obtained from the battery-solar system. The battery-solar system functions as follows: the solar panel captures solar energy, converting it into electricity, which is then regulated through an MPPT (Maximum Power Point Tracking) module to charge the battery. This stored electricity powers all other components of the CPS. Output signals, such as the energy level of the solar-battery system are broadcasted to the blockchain, serving as input signals for calculating the allowable rental time for no1s1. This calculated time is automatically updated and displayed on the user interface, enabling users to select their preferred duration. For instance, when the energy level reaches 80%, users have the option to rent the space for more than 2 hours. However, during testing phases, this process defaults to few choices due to time constraints within the testing environment.

The second feedback loop set governs the process of physically entering the space of no1s1, as shown in Figure 1(a). In these loops, human actions and identity are verified through a camera and an Ultrasonic Distance Sensor (UDS). After selecting a desirable usage duration from the last step, the user needs to make a deposit payment to the house which is default to 1 test ETH in many testing cases. The success of the payment will grant the user an QR code, which is calculated cryptographically with the input of username, location and wallet address. The QR code is required to be shown to the camera of the cabin and the cabin will then quest the user information from the blockchain network to check the payment status of the user. The smart door unlocks when payment is confirmed and then the UDS starts to detect the presence of humans through measuring the distance to the door. The UDS mechanism operates by calculating the time delay of the echo produced when a pulsed ultrasound wave encounters an object. Alongside UDS signals, the state of the smart lock—whether it is locked or released—is also considered as input to determine if access can be granted to the user. Subsequently, when the user utilises the space, the signal state changes accordingly. For instance, access cannot be granted when the house is already occupied, as indicated by the signals.

## 6   No1s1 Governance

After outlining the technical configuration, and control and feedback loops of no1s1, the next consideration of cybernetic principles is how such an autonomous CPS agent can be governed in a decentralised manner. To tackle this, we synthesise and apply theories from first and second-order cybernetics. As defined, first-order cybernetic systems are observed and controlled via sensors and actuators that create feedback loops. Extending beyond common CPS feedback loop architecture for intelligent buildings, no1s1 feedback loops insert a blockchain network as an intermediary step where important states of the CPS are stored in a decentralised manner, akin to a Knowledge Commons Institution (KCI) (Figure 2 (b)).

Second-order cybernetics acknowledges that third-party observers of a system are outside of that system, thus creating an overarching feedback loop between the observer and the observed. The purpose of second-order cybernetics is to create circulatory, and thus reinforce the core cybernetic principles of information and control [21]. In no1s1, the blockchain ledger serves as a secure single source of truth of the system's status, ensuring relevant parties (explored in more detail below) have access to the necessary information to inform meta-policy parameters in the smart contracts, such as governance rights. For example, the access price is a parameter that also signifies the governance. as the price can be determined through a predefined governance mechanism encoded in smart contracts. This encoded algorithmic policy loop with adaptive and changeable parameters in the CPS forms the basic building block of the KCI [34]. In this context, the KCI functions as an observer, informing governance—effectively serving as a meta-controller that supervises individual controllers and interacts with sensors/actuators within the system. The no1s1 second-order governance feedback loops, inspired by crypto-economic governance feedback loops [35], encompass micro, meso, and macro layers. The micro layer reflects the behaviour of both human and machine agents, while the macro layer amplifies the system's purpose and phenomena. The meso layer is the governance layer, where policies and rules are formulated to align agent-level behaviour with the system-level purpose. Such an institution entails a transparent knowledge commons, a consensus mechanism, and encoded governance rules and policies for viable and sustainable organising [17]. The section that follows explores some of the considerations in governing a decentralised autonomous CPS.

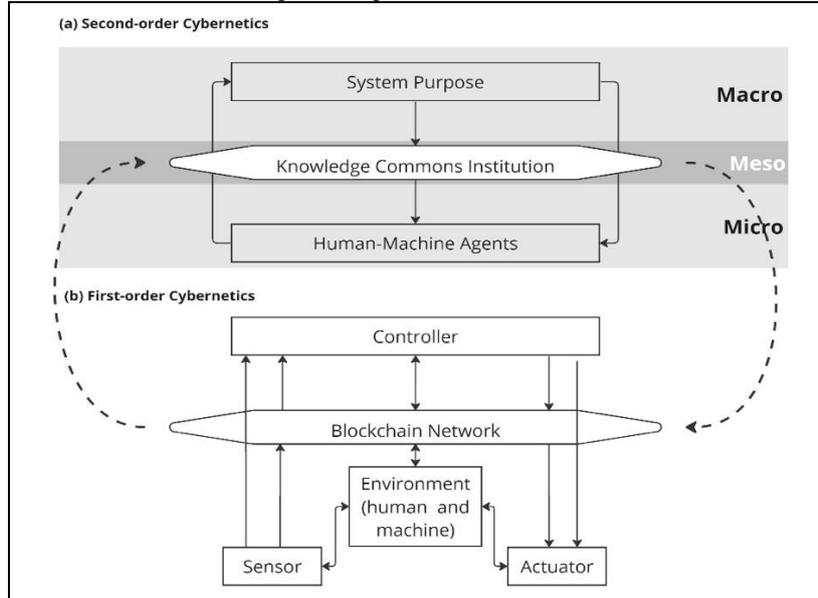

**Fig. 2.** no1s1 governance from first order to second order cybernetics

## 7 Considerations in the Decentralised Governance of CPSs

Governance of a CPS is predicated on identifying the "governance surface", as the set of parameters through which a system's boundaries or rule-set can be modified [18, 35], and then mapping the parameters in which various stakeholders can contribute in particular ways.
The governance surface of no1s1 is what exists beyond the physical confines of the cabin, as the second order control loops that enable governance of automation of the physical system (as illustrated in Figure 2). While the previous section outlined the governance system, numerous questions pertaining to governance remain unanswered [16]. The first consideration is what stakeholders are involved in the governance of a CPS, what parameters do they have permission to govern, and how are governance rights granted? For no1s1, the operation of the physical cabin itself is autonomous, meaning it can manage occupancy and payments, according to the control loop described above. Yet, meta parameters still need to be determined further as to the overarching objectives of the system, as well as what the system optimised for.

The underlying logic behind the governance and economics of the system needs to be determined. Identifying the stakeholders who should have governance rights may include users, operators, developers, and potentially other affected communities. For example, governance rights to no1s1 could be based on capital, by awarding steering and decision-making influence to those that built that system, in order to make a return on their investment. In contrast, it could be a user-based system, where membership and governance authority is allocated based on people's use of the house. Another option could be a labour-based system, whereby those that monitor and maintain the function of the house are granted a say in setting its governance surface. Quite likely, no1s1 may benefit from a mix of all three of these approaches, with various levels of governance rights granted to each category of stakeholder. The types of governance rights may also differ among different stakeholder groups. For example, these could range from voting rights, the right to propose changes or updates to the system, or the right to veto certain decisions.

From here, stakeholders need a clear remit in governing a house that is financially autonomous. If no1s1 is based on a raw capitalist system of maximising revenue, it will optimise to maximise occupancy and rates, and minimise costs of cleaning and maintenance. Yet, in this scenario, the CPS lacks broader social and environmental context or values, and risks becoming highlight efficient but not in line with its intended purpose (known as instrumental convergence, or "paperclip maximisation"[36]). The necessity of human governors in the system is to bring broader context to the system to help set the parameters of the governance surface in ways that are ultimately useful and in line with its purpose. For example, in relation to no1s1, this might include adding a sustainability goal to harmonise with the environment, or metrics in support of socio-economic inclusivity, as a factor external to, and not in the financial interests of a house. Additionally, when extending beyond the scale of a cabin, the governance rights would also be largely influenced by the archetype of the space. For example, a public space might be way better-off with participatory governance, while privately owned space might have a smaller stakeholder circle and can be gated to whom can participate into the governance processes.

A second consideration in the decentralised governance of CPSs is the actual allocation of governance rights and the structuring of governance processes. These considerations involve establishing who holds decision-making powers, and how through what processes, mechanisms, and other infrastructure decisions are made. Deciding on the model of decision-making to be used, which could include direct democracy (whereby every stakeholder votes on every decision), representative democracy (whereby stakeholders elect representatives to make decisions), or liquid democracy (a hybrid allowing stakeholders to either vote directly or delegate their vote to a representative). Furthermore, mechanisms for achieving consensus among stakeholders may vary from simple majority voting, to more complex mechanisms like quadratic voting (a popular mechanism in some blockchain communities such as the "Gitcoin" crowd-funding platform, where the cost of each additional vote increases quadratically [37]) or other consensus algorithms utilised in blockchain governance. How these governance processes will remain adaptive to incorporate necessary changes and improvements (including data management processes and security), as well as who and how they will be held accountable, also requires careful consideration.

The third consideration involved different phases of the house. A house can have a longer lifespan than a human, sometimes hundreds of years. On top of this, the lifecycle of a house involves different phases, such as design, planning, construction, operation, maintenance, and end-of-life (demolition or reconstruction). This indicates a changing constituency of stakeholders over time, possibly with different incentives and considerations in different phases of the house CPS lifecycle. For example, to even start the planning of the house, funding needs to be secured. This initial funding mechanism might differ greatly from budgeting for and maintenance phase, therefore requiring a different decision-making scheme. The requirement in level of collaboration, type of interaction, and speed of operation also varies greatly across phases. Thus, the governance mechanism could benefit from an adaptive framework, rather than a static or rigid one.

# 8 Challenges in Decentralised Governance of Autonomous Physical Spaces

CPSs also introduce a range of challenges to consider, that can be limitations on effectiveness if not considered and managed. This paper opens up several avenues for further research, including security, scalability, and evolution. As a feature of CPS, real-time data processing enables immediate responses to changes in the environment or occupants' behaviours and needs. Yet, CPS reliance on networked computer infrastructure also adds the requirement of robust cybersecurity measures to protect against unauthorised data access to ensure data privacy [38]. Numerous organisations and security standards bodies have issued specific guidelines for integrating cybersecurity features into CPS that include functional concerns regarding the systems sensing, actuating, and control, as well as business, human, data, and lifecycle of CPS components. Identification and careful considerations of these factors is essential for both interoperability between CPSs, as well as aligning decentralised governance processes with existing legal and regulatory frameworks for compliance. Further complicating this requirement is that legal frameworks may be designed around more centralised models of governance, which is already a challenge in the decentralised governance space. Other areas for further research include the scalability of governance structures and mechanisms for CPS according to their lifecycle, to ensure they remain effective as they expand and their user base grows. Additionally, exploring the decentralised governance of computing, especially in the context of AI, presents significant opportunities to understand how approaches to decentralised governance of physical domains can be more effective. Finally, future research could also explore the trajectory of initiatives like no1s1, assessing their potential evolution into smart homes and cities, and their role in environmental impact monitoring and reporting to contribute to more sustainable urban development.

# 9 Conclusion

This paper has presented a novel exploration of the considerations in the decentralised governance of autonomous, physical spaces to achieve self-agency. By demonstrating the practical aspects of the cybernetic principles of decentralised governance of Cyber-Physical Systems, as evidenced in the no1s1 case study, alongside delineating the challenges and limitations associated with the deployment and governance of CPS, this research serves as a critical foundation for future governance models in scalable, blockchain-enabled Decentralised Physical Infrastructure Networks. This investigation not only contributes to the fields of infrastructure studies and Cyber-Physical Systems engineering but also enriches the discourse on decentralised governance and the multifaceted nature of autonomy in technological and societal systems, offering insights for future research and application in decentralised governance and management of autonomous physical spaces.

## 11  Appendix

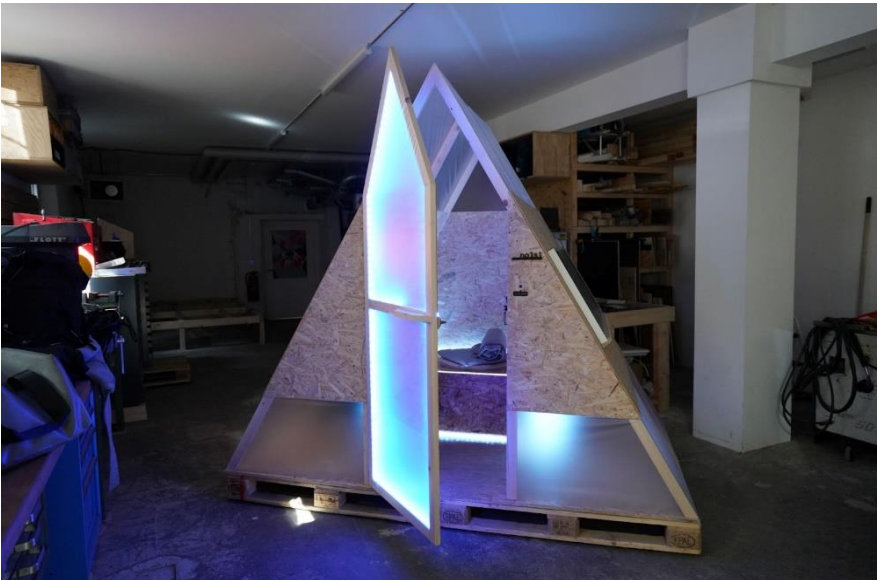

**App. 1.** no1s1 prototype